\documentclass{article}
\usepackage{graphicx}
\usepackage{amsmath, amssymb}
\usepackage{booktabs}
\usepackage{geometry}
\usepackage{caption}
\usepackage{subcaption}
\usepackage{filecontents}
\usepackage{biblatex}
\title{Futility Analysis under Scrutiny}
\author{Rui Jin, Cai Wu, Peter Mesenbrink}

\date{July 2025}

\begin{document}

\maketitle

\begin{abstract}
This paper investigates the robustness of futility analyses in clinical trials when interim analysis population deviates from the target population. We demonstrate how population shifts can distort early stopping decisions and propose post-stratification strategies to mitigate these effects. Simulation studies illustrate the impact of subgroup imbalances and the effectiveness of naive, model-based, and hybrid post-stratification methods. We also introduce a permutation-based screening test for identifying variables contributing to population heterogeneity. Our findings support the integration of post-stratification adjustments using all available baseline data at the interim analysis to enhance the validity and integrity of futility decisions.
\end{abstract}

\section{Introduction}\label{sec1}
Futility analysis plays a pivotal role in the interim monitoring of clinical trials, offering a mechanism to terminate studies early when accumulating evidence suggests that the investigational treatment is unlikely to achieve its intended clinical benefit~\cite{demets2006}. This approach not only conserves resources and protects participants from unnecessary exposure under inferior treatments but also enhances the ethical and operational efficiency of drug development. A wide range of statistical frameworks—spanning frequentist group sequential designs to Bayesian predictive and posterior probability-based methods—have been developed to guide futility decisions, reviews can be found in \cite{lachin2005} and \cite{snapinn2006}. These methods are typically evaluated under idealised conditions, assuming that the interim analysis (IA) dataset is a representative sample of the target population. 

However, in practice, this assumption of representativeness is often violated. Variability in site-level recruitment, temporal enrolment patterns, and stratified randomisation can lead to interim populations that differ systematically from the target population. Such deviations introduce what we term extrinsic error—biases arising not from sampling variability but from structural differences in subgroup composition. These errors can distort the operating characteristics of futility rules, inflating type I errors (premature termination of effective treatments) or type II errors (failure to stop ineffective treatments). Despite its practical implications, the impact of population shifts on futility decision-making has received limited attention in the statistical literature.

This paper addresses this methodological gap by systematically evaluating the robustness of futility rules under violations of the representativeness assumption. Through simulation studies, we demonstrate how even modest deviations in subgroup composition at IA can substantially alter the probability of early stopping. To mitigate these effects, we explore a suite of post-stratification~\cite{holt1979} strategies—including naive re-weighting, model-based adjustments, and a novel hybrid approach—that aim to restore the intended operating characteristics of futility analyses in the presence of population heterogeneity.

The remainder of the paper is organised as follows. Section~\ref{sec2} introduces the simulation framework and illustrates the sensitivity of futility rules to subgroup composition shifts. Section~\ref{sec3} presents post-stratification methods, including naive, model-based, and hybrid estimators, and evaluates their performance. Section~\ref{sec4} extends these methods to settings with incomplete baseline data. Section~\ref{sec5} proposes a permutation-based screening test to identify variables contributing to population heterogeneity. Section~\ref{sec6} concludes with practical recommendations.

\section{Robustness of a Futility Analysis}\label{sec2}
Simulation-based evaluations of futility rules typically begin with the generation of complete patient-level datasets, from which subsets are drawn to assess operating characteristics of futility rules at various information fractions. This framework implicitly assumes that the interim analysis dataset constitutes a representative sample of the full trial population—an assumption equivalent to data being missing completely at random (MCAR) for patients not yet accrued. While convenient, this assumption is often unrealistic in practice. In this section, we investigate the robustness of futility rules when the representativeness assumption is violated.

Stratified randomization is commonly employed at trial initiation to ensure balance across key prognostic or confounding variables (e.g., disease subtype, age, sex). However, due to site-level enrollment variability and temporal recruitment dynamics, the IA population may not reflect the intended stratification.  For instance, early-enrolling sites may predominantly recruit patients from a specific demographic or disease subtype, leading to overrepresentation of certain strata and underrepresentation of others. This imbalance can result in the IA population deviating from the full population in terms of key covariates, introducing unanticipated heterogeneity. Such discrepancies may distort treatment effect estimates and compromise the validity of early stopping decisions. To investigate this issue, we examine the performance of futility rules under scenarios where the IA population is not representative of the overall trial population.

To evaluate robustness of futility rules, we evaluate how population shifts influence the probability of incorrect decisions—specifically, type I errors (premature termination of an efficacious treatment) and type II errors (failure to terminate an ineffective treatment). We further decompose these errors into two components: intrinsic and extrinsic. Intrinsic error arises from statistical uncertainty due to limited sample size and can typically be mitigated by increasing the IA sample size. In contrast, extrinsic error stems from systematic differences between the IA and full populations (e.g., imbalances in stratification factors) and cannot be addressed through increasing sample size alone. Our focus is on the contribution of representativeness violations to extrinsic error and the extent to which such violations distort the operating characteristics of futility rules.

We conducted a simulation study to illustrate how unanticipated population heterogeneity can compromise the operating characteristics of futility rules. Consider a randomized controlled trial with a planned enrollment of 300 patients, randomized 1:1 to treatment and placebo arms. The primary endpoint is a binary outcome assessed at study completion. The target population comprises two subgroups in a 60:40 ratio, each with a baseline response rate of 30\%. Under treatment, the assumed response rates for the subgroups are 40\% and 65\%, respectively, yielding an overall treatment response rate of 50\%.

At the time of IA, planned at 40\% of participants completed the primay endpoint assessment, the observed subgroup proportions deviate from the planned 60:40 distribution, introducing a population shift. This deviation violates the assumption of representativeness at IA and introduces heterogeneity that can bias the estimation of the treatment effect. The futility rule under consideration is based on the posterior probability that the treatment effect exceeds a clinically meaningful threshold of 20\%. This probability is computed using a beta-binomial model with a uniform prior. If the posterior probability is less than 0.1, the trial is deemed futile and may be terminated early. Under the true population structure, the posterior probability would support continuation. However, due to the altered subgroup composition at IA, the estimated treatment effect may be attenuated, increasing the risk of a false futility declaration.

\begin{figure*}[h!]
    \centerline{\includegraphics[width=\textwidth,height=17pc]{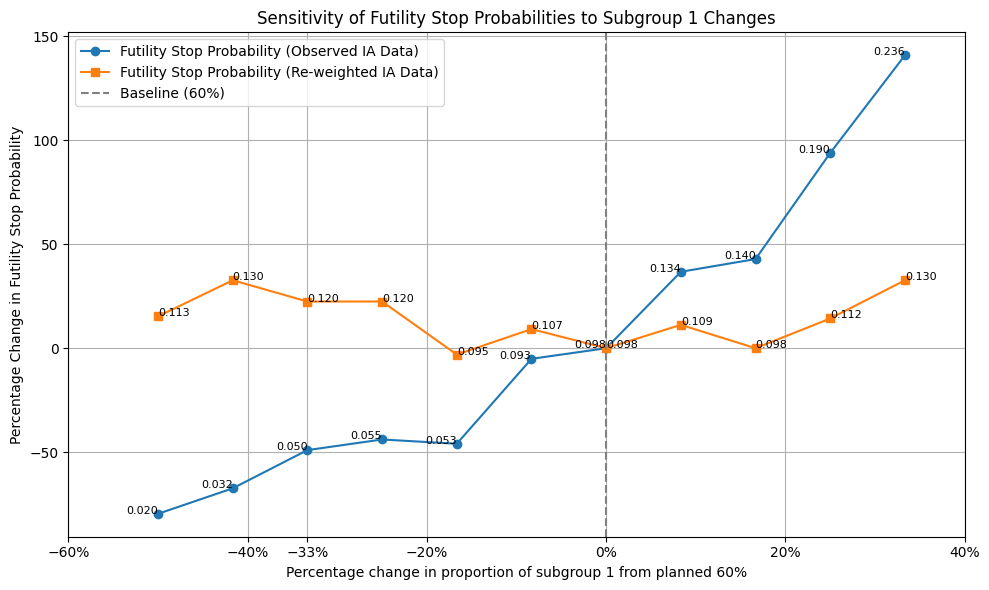}}
    \caption{Impact of subgroup proportion shifts on futility stop probability (posterior probability).}
    \label{ex1}
\end{figure*}

Figure~\ref{ex1} illustrates the sensitivity of the futility stop probability to deviations in subgroup composition. Each point is the average of 1000 repetitions. The x-axis represents the percentage change in the proportion of subgroup 1 from the planned 60\% in the IA dataset, while the y-axis shows the corresponding percentage change in the probability of stopping for futility. The blue curve depicts how the futility stop probability varies with different subgroup 1 proportions. Notably, when the IA dataset contains subgroups in a 70:30 ratio (a 16.7\% shift), the futility stop probability increases by approximately 42.8\%. This example underscores the vulnerability of posterior probability-based futility rules to distributional shifts and highlights the importance of accounting for population heterogeneity in interim decision-making.

\begin{figure*}[h!]
        \centerline{\includegraphics[width=\textwidth,height=17pc]{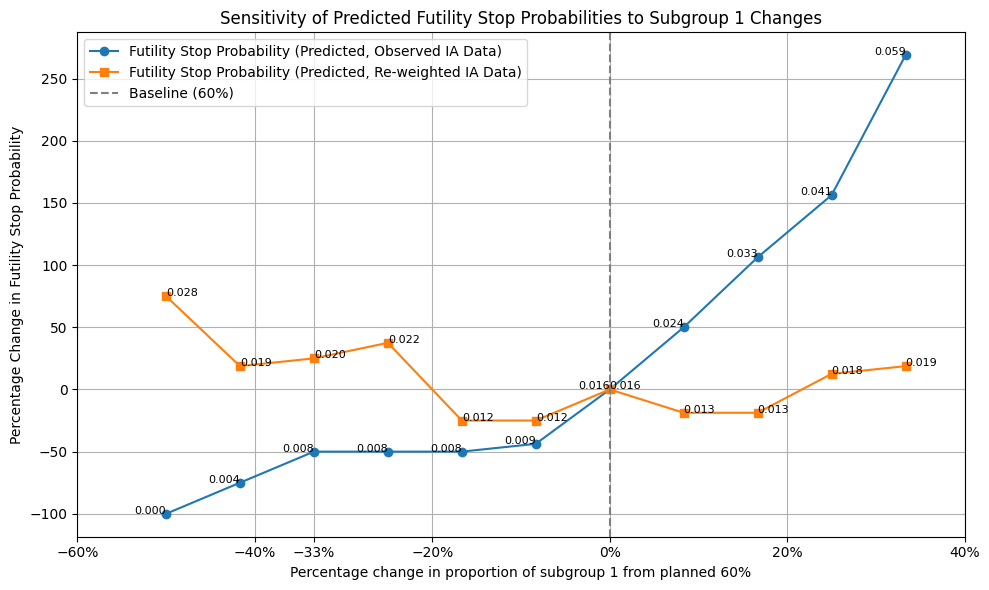}}
    \caption{Impact of subgroup proportion shifts on futility stop probability (predictive probability).}
    \label{ex1_pred}
\end{figure*}

To mitigate the impact of such imbalances, we propose a re-weighting approach that restores the intended subgroup distribution.  Specifically, each subgroup is assigned a weight equal to the ratio of its true population proportion to its observed sample proportion at interim. This adjustment compensates for deviations in subgroup representation and corrects the bias introduced by population shifts. Applying the futility rule to the re-weighted sample mean successfully restores the original error rates, as shown by the yellow curve in Figure~\ref{ex1}. This result demonstrates that appropriate weighting can effectively counteract the effects of population shifts at interim analysis and preserve the intended operating characteristics of the futility monitoring rule. For a predictive probability based futility rule, similar results can be achieved. Comparable improvements are observed when the same re-weighting strategy is applied to predictive probability-based futility rules (Figure~\ref{ex1_pred}).

We further note that the re-weighting approach described above is mathematically equivalent to a post-stratification method based on observed subgroup mean response rates.  Let $n_1$, $n_2$ denote the observed sample sizes for the subgroup1 and subgroup2 subgroups, respectively, and let  $p_1$, $p_2$ represent their true population proportions. The re-weighting method adjusts each subgroup’s contribution by the ratio of its true population proportion to its observed sample proportion. Specifically, it is
\[
\text{Adjusted Average} = \frac{1}{n_1 + n_2} \left(  p_1 \cdot \frac{\bar{r}_1 \cdot n_1}{\frac{n_1}{n_1 + n_2}} +  p_2 \cdot \frac{\bar{r}_2 \cdot n_2}{\frac{n_2}{n_1 + n_2}} \right) =  p_1 \cdot \bar{r}_1 + p_2 \cdot \bar{r}_2 \,,
\]
where $\bar{r}_1$ and $\bar{r}_2$ are observed average response rates for subgroup1 and subgroup2, respectively.
This formula represents a special case of the widely used post-stratification method, which will be introduced in detail in the next section.

In the example presented in this section, we made several simplifying assumptions. First, the true population proportions for each subgroup are known. Second, observations are available for all subgroups under consideration. Third, we assume that the variables responsible for population shifts in the IA population are known and correctly identified. In practice, however, these assumptions likely will not hold. For instance, population proportions may be uncertain or estimated with error, and some subgroups may be underrepresented or entirely missing from the IA dataset. Additionally, the factors driving population shifts may be unknown or only partially observed. These practical limitations and their implications for the validity of post-stratification adjustments and futility rule performance will be discussed in later sections of this paper.

\section{Post stratification}\label{sec3}

Post-stratification is a statistical adjustment used to improve the representativeness of sample-based estimates when the observed sample distribution deviates from known population characteristics, reference of this topic can be found in \cite{little1993}, \cite{holt1979}, \cite{zhang2000} and \cite{miratrix2013}. After data collection, the sample is partitioned into \( K \) strata based on auxiliary variables (e.g., age, sex, region), and weights are applied to align the sample with the true population structure. Let \( \hat{Y}_k \) denote the estimated sample mean within stratum \( k \), \( n_k \) the number of observations in stratum \( k \), and \( p_k \) the known population proportion for that stratum, where \( k = 1, \dots, K \). The post-stratified estimator of the overall mean is then given by:

\begin{equation}
    \bar{Y}_{\text{post}} = \sum_{k=1}^{K} p_k \hat{Y}_k \,.
\end{equation}

This estimator will replace the unweighted sample mean with a weighted average that reflects the true population composition. Post-stratification is particularly useful in survey sampling and potentially in interim analyses of clinical trials, where subgroup representation may be unbalanced due to enrollment variability. By incorporating known population margins, it improves the accuracy and validity of statistical inference without altering the original sampling design.

Various post-stratification methods have been developed, differing primarily in the estimation of stratum-specific means. The most straightforward one is to use the observed sample mean within each stratum, which will be referred as the naive approach. However, this method can perform poorly when certain strata contain only a small number of observations, leading to high variability. In extreme cases—such as when a stratum has no observations at all—the method breaks down entirely. To address these limitations, model-based post-stratification approaches have been proposed. These methods borrow strength across strata by incorporating auxiliary information or fitting statistical models, thereby improving estimation stability and enabling inference even in the presence of sparse or missing strata.

Since futility rules rely primarily on interim estimates derived from observed sample means, it is essential to understand how post-stratification adjustments influence these estimates under varying degrees of population imbalance. To this end, the simulation studies presented in this section focus on the distributional behavior of both unadjusted sample means and post-stratified means across a range of subgroup compositions. Importantly, these simulations are designed to isolate the effect of post-stratification by excluding predictive or posterior probability calculations, thereby providing a direct assessment of how weighting strategies impact the stability and accuracy of interim estimates used in futility decision-making.

\subsection{Model based post-stratification example: Bayesian multilevel regression}

Bayesian multilevel regression~\cite{park2004}, also known as hierarchical modeling, is a flexible statistical framework for analyzing data with nested or grouped structures. It enables simultaneous estimation of group-specific parameters while borrowing strength across groups through shared hyperparameters, improving estimation stability—especially in settings with sparse or unbalanced data.

To illustrate, suppose we have \( G \) groups, and for each group \( i \), we observe \( n_i \) data points \( y_{ij} \), where:

\[
y_{ij} \sim \mathcal{N}(\mu_i, \sigma^2), \quad j = 1, \dots, n_i
\]

The group-specific means \( \mu_i \) are assumed to follow a common prior distribution:

\[
\mu_i \sim \mathcal{N}(\mu, \tau^2)
\]

This defines a two-level hierarchical model: the first level models the data within each group, and the second level models the distribution of group means across the population. Given the data, the posterior distribution of each group mean \( \mu_i \) is:

\[
\mu_i \mid \bar{y}_i \sim \mathcal{N} \left( \frac{\tau^2}{\tau^2 + \sigma^2 / n_i} \bar{y}_i + \frac{\sigma^2 / n_i}{\tau^2 + \sigma^2 / n_i} \mu,\ \left( \frac{1}{\tau^2} + \frac{n_i}{\sigma^2} \right)^{-1} \right)
\]

The posterior mean is a weighted average of the group-specific sample mean \( \bar{y}_i \) (the no-pooling estimate) and the overall mean \( \mu \) (the complete-pooling estimate):

\[
\hat{\mu}_i^{\text{partial}} = w_i \bar{y}_i + (1 - w_i) \mu
\]

where the shrinkage weight \( w_i \) is given by:

\[
w_i = \frac{\tau^2}{\tau^2 + \sigma^2 / n_i}
\]

Here, \( \tau^2 \) represents the between-group variance, \( \sigma^2 \) the within-group variance, and \( n_i \) the sample size for group \( i \). This formulation allows the model to adaptively shrink group estimates toward the overall mean, with the degree of shrinkage determined by the relative amount of information available within each group. The following plots illustrates this partial pooling effect, highlighting how Bayesian multilevel regression balances between no pooling and complete pooling to produce more stable and reliable estimates.

\begin{figure*}[h!]
       \centerline{\includegraphics[width=\textwidth,height=17pc]{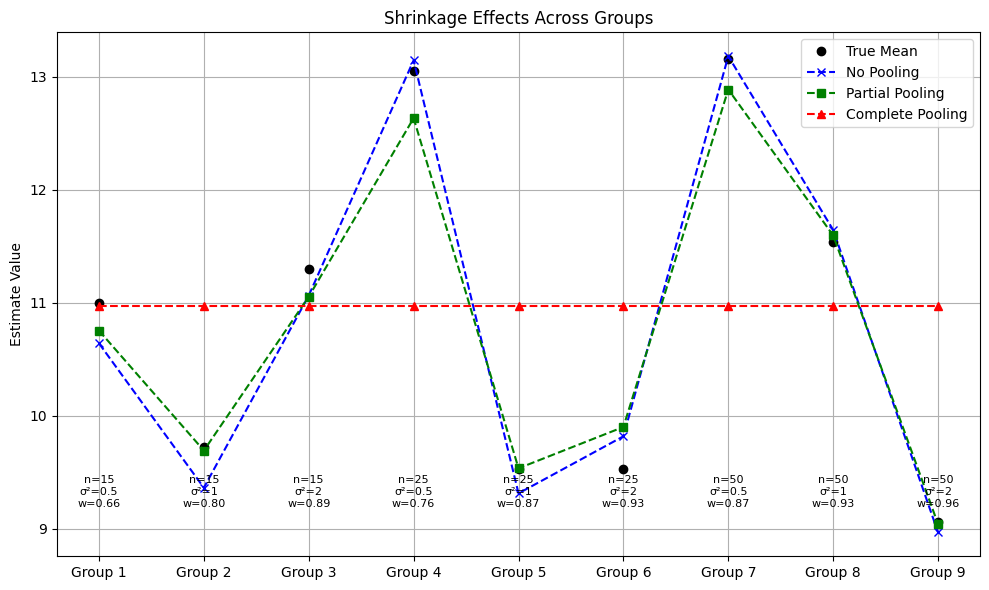}}
    \caption{Impact of $n$, $\sigma^2$ on the shrinkage weight}
    \label{pooling_1}
\end{figure*}

Figure~\ref{pooling_1} illustrates the shrinkage effect in a Bayesian multilevel model, where group-level estimates are pulled toward the overall mean. The degree of shrinkage is governed by both the number of observations within each group and the assumed between-group variability. Specifically, groups with fewer observations or lower assumed variability experience greater shrinkage toward the global mean.

\begin{figure*}[h!]
       \centerline{\includegraphics[width=\textwidth,height=17pc]{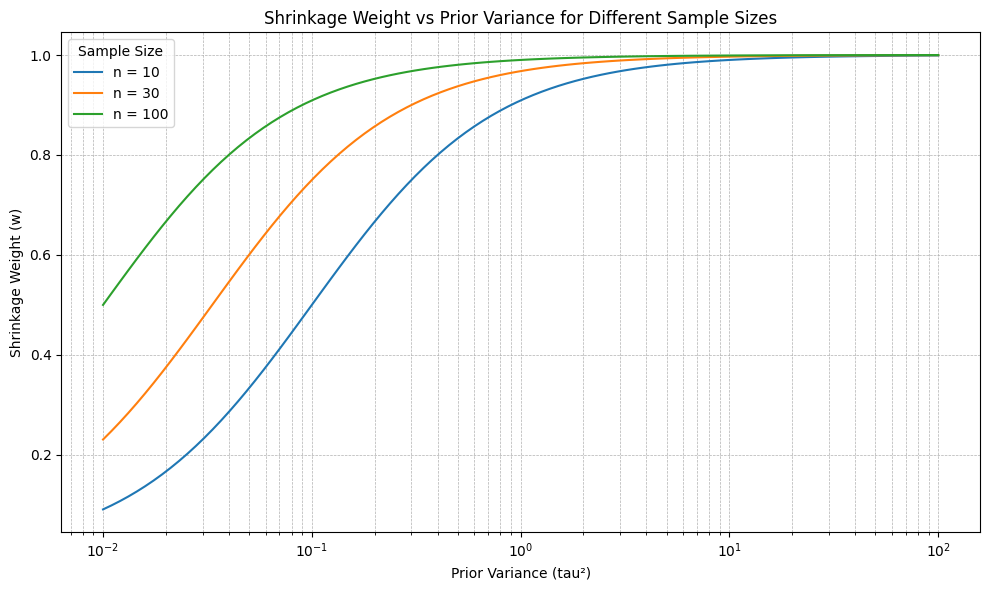}}
    \caption{Impact of $n$, $\tau^2$ on the shrinkage weight}
    \label{pooling_2}
\end{figure*}

Figure~\ref{pooling_2} further explores how the shrinkage effect varies with the prior variance $\tau^2$ in the hierarchical model across different group sample sizes. When $\tau^2$ is small, the model assumes minimal heterogeneity between groups, resulting in stronger shrinkage. As the sample size within a group increases, the shrinkage weight approaches one, indicating that the model increasingly relies on the group-specific estimate rather than the overall mean. This behavior reflects the adaptive nature of Bayesian hierarchical models in balancing information across levels of the hierarchy.

\subsection{Impact of model misspecification on model-based post-stratification}

Model-based post startification methods are widely used for estimating population-level quantities from non-representative samples. Their accuracy, however, critically depend on the correct specification of the underlying model~\cite{black2011}. When the model is misspecified—such as by omitting key interactions or misrepresenting group-level effects—it can introduce systematic bias. This bias is particularly problematic in strata with large post-stratification weights, where even small errors in prediction can disproportionately influence the overall estimate. The following example illustrates how such misspecification can distort results in model-based post-stratification approaches.

Consider a randomized controlled trial with a planned enrollment of 600 patients, randomized in a 1:1 ratio to treatment and placebo arms. The primary endpoint is a continuous outcome assessed at study completion. The target population comprises two subgroups in a 60:40 ratio, with a baseline average score of 40 and 20, respectively. Under treatment, the assumed subgroup-specific effects are 7 and 3 units, respectively, yielding an overall treatment effect of 5.2 units. To account for site-level variability, a random site effect with four observed levels is incorporated into the response variable. This site effect is modeled as a normally distributed random variable with mean zero and standard deviation 2.5.

At the time of interim analysis, conducted after 40\% of participants have completed the primary endpoint assessment, the observed subgroup proportions deviate from the planned 60:40 distribution and instead reflect a 50:50 composition. This population shift violates the assumption of representativeness at IA and introduces heterogeneity that may bias the estimation of the treatment effect.

\begin{figure*}[h!]
           \centerline{\includegraphics[width=\textwidth,height=17pc]{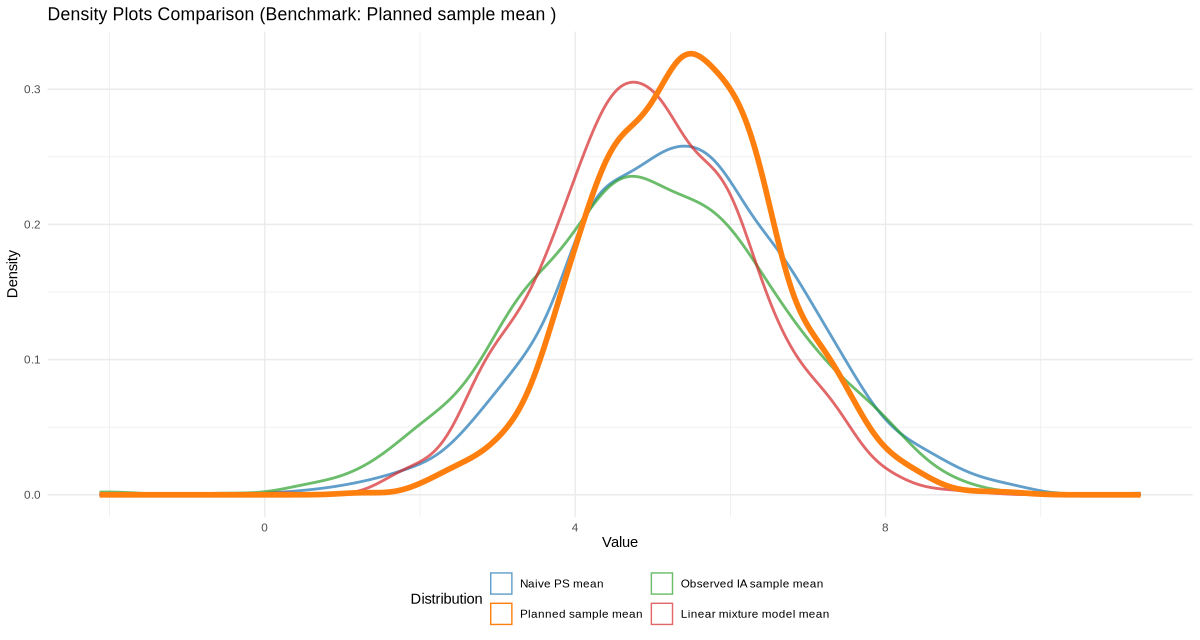}}
    \caption{Empirical distributions of estimated mean difference}
    \label{ex2_1}
\end{figure*}

To investigate the impact of this shift, we compare four estimation methods for the sample mean based on IA data. Empirical distributions of the estimated treatment effect were obtained from 1,000 simulation replicates, which are shown in figure~\ref{ex2_1}. The benchmark method (orange) reflects estimation using 40\% of data sampled from the planned population distribution, serving as a reference. The observed IA sample mean (green) is computed directly from the shifted population. The naive post-stratification method (blue) adjusts the IA sample using known population proportions. Finally, the misspecified linear mixed effects model (red) assumes homogeneous treatment effects across subgroups.

The results indicate that both post-stratified methods yield estimates with reduced variance compared to the unadjusted IA sample mean. However, the model-based approach suffers from bias due to incorrect assumptions about treatment homogeneity, resulting in inferior performance relative to the naive post-stratification method. These findings underscore the importance of accounting for subgroup-specific effects and population composition when estimating treatment effects at interim analysis.

\subsubsection{Hybrid post stratification method}

To address the limitations of model-based post-stratification under misspecification, we propose a hybrid post-stratification approach that integrates naive and model-based estimation. For the $k$th stratum, if the number of observed units exceeds a predefined threshold, we directly use the observed sample mean as $\hat{Y}_k$. In contrast, for strata with limited or no observations, we substitute the model-predicted mean as $\hat{Y}_k$ in the post-stratification process. This hybrid approach leverages empirical data where it is reliable and defers to model-based estimates where data are sparse, aiming to reduce bias caused by model misspecification and improve robustness.

\begin{figure*}[h!]
    \centerline{\includegraphics[width=\textwidth,height=17pc]{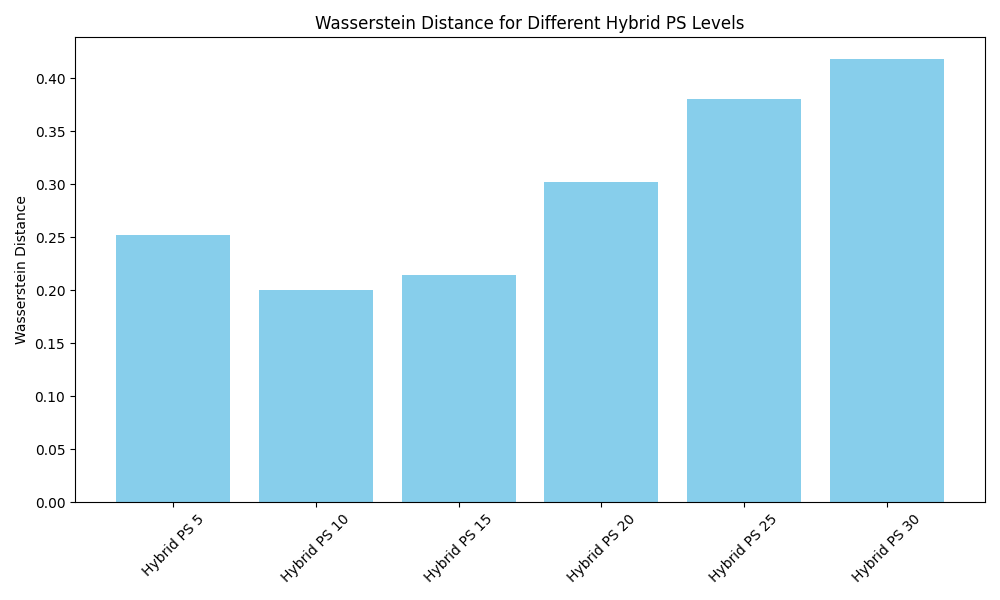}}
    \caption{Wasserstein distance between empirical distributions of hybrid estimates to the benchmark}
    \label{ex2_3}
\end{figure*}

The choice of threshold for switching between empirical and model-based estimates is data-dependent and can be optimized through simulation. We evaluated the performance of the hybrid estimator across a range of cutoff values using the Wasserstein $L_1$ distance between the empirical distribution of hybrid estimates and a benchmark distribution derived from a representative interim sample. As shown in Figure~\ref{ex2_3}, a cutoff of 10 observations yielded a favorable trade-off between bias and variance, avoiding overfitting while maintaining proximity to the benchmark. This optimality is supported by the fact that the Wasserstein distance quantifies the minimal cost of transforming one distribution into another, making it a robust metric for assessing distributional similarity. A smaller Wasserstein distance indicates that the hybrid estimator closely approximates the benchmark distribution, reflecting both accuracy and stability. The cutoff of 10 thus minimizes estimation error by leveraging empirical data when sufficient observations are available, while defaulting to model-based estimates in sparse regions to reduce variance. This balance ensures that the estimator remains both data-adaptive and generalizable.

\begin{figure*}[h!]
      \centerline{\includegraphics[width=\textwidth,height=17pc]{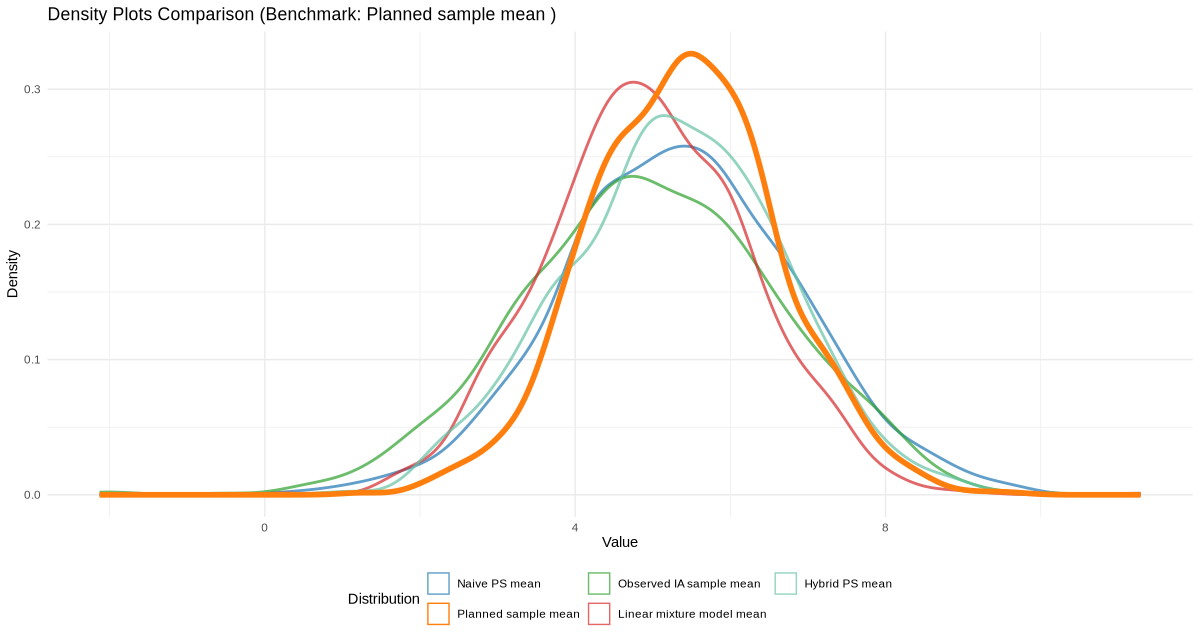}}
    \caption{Empirical distributions of estimated mean difference (including the hybrid approach)}
    \label{ex2_2}
\end{figure*}

To further assess the utility of the hybrid approach, we applied it to the previously discussed scenario involving model misspecification. Figure~\ref{ex2_2} presents the comparative empirical distributions. The hybrid estimator provide a better approximation of the benchmark distribution, outperforming the model-based estimator, which suffers from bias due to incorrect assumptions. Compared to the naive post-stratification method, the hybrid approach exhibits reduced variability, demonstrating its ability to adaptively combine the strengths of both empirical and model-based strategies.

\section{Post stratification based on estimated proportions for stratum}\label{sec4}

In practice, full baseline data for all participants may not be available at the time of interim analysis, except in studies with long follow-up periods for primary endpoint collection. In most cases, futility analyses are designed to allow early termination of trials when there is strong statistical evidence indicating a lack of treatment effect.

When full baseline data are unavailable for estimating stratum proportions, we propose using the available data to approximate these proportions. For example, variables such as disease subtype and gender may already be well-studied in the datasets of early trials. If such data are not available, we can still rely on the observed baseline characteristics to estimate stratum proportions and identify variables contributing to unexpected population heterogeneity.

To assess the uncertainty associated with the estimated post-stratified sample means under incomplete baseline data, simulation studies are recommended. These simulations can help evaluate the robustness of  post-stratification approaches under varying degrees of data availability.

We present a simulation study based on the scenario described in Section~2. We fix the observed IA subgroup composition at 70:30. As previously discussed, such deviation from the planned distribution can inflate the probability of erroneously stopping the trial. In this setting, we investigate the impact of estimating population proportions at IA using partially available baseline information. Specifically, we assume that the true subgroup proportions are unknown and must be inferred from baseline data available at the time of IA.

\begin{figure*}[h!]
    \centerline{\includegraphics[width=\textwidth,height=17pc]{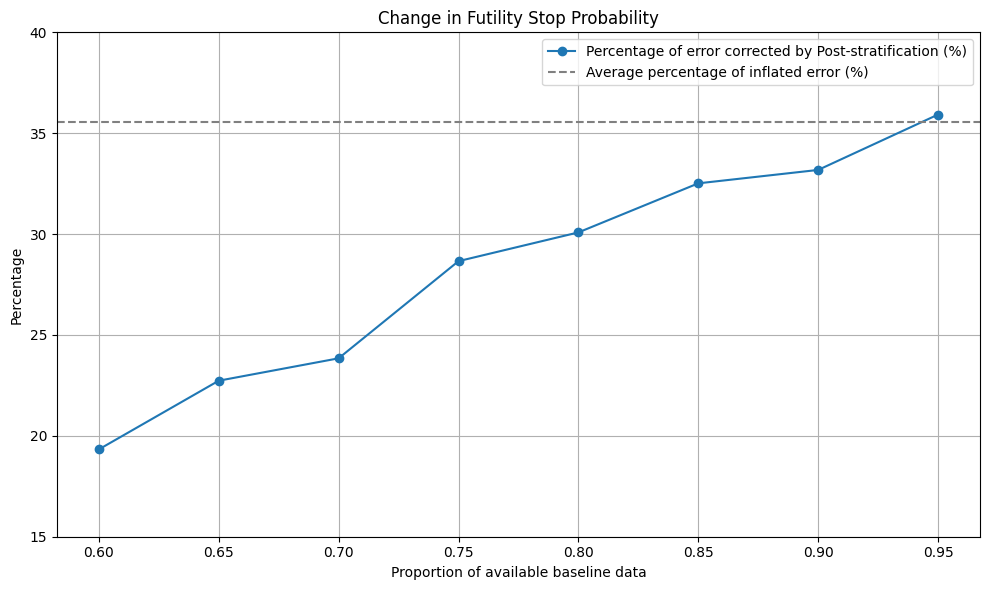}}
    \caption{Amount of correction for the inflated error via post stratification}
    \label{ex3}
\end{figure*}

To evaluate the sensitivity of post-stratification, we vary the proportion of available baseline data used to estimate subgroup proportions from 60\% to 95\%. For each level of data availability, we first estimate the subgroup proportions and then compute futility probabilities based on the corresponding post-stratified mean. Results are presented in figure~\ref{ex3} where each point is based on 10000 replicated simulations. On average, the error inflation in futility probability due to directly using IA data is 35.5\%. The plot shows that post-stratification using subgroup proportions estimated from 60\% of baseline data can still correct approximately 19\% of the inflated error. These findings demonstrate that even when subgroup proportions are estimated from incomplete baseline data, the post-stratified mean-based futility rule can substantially reduce the error rate compared to unadjusted methods. This finding suggests that post-stratification based on all available baseline data remains beneficial in interim analyses, even when true population proportions are not fully known. 

\begin{figure*}[h!]
  \centerline{\includegraphics[width=\textwidth,height=17pc]{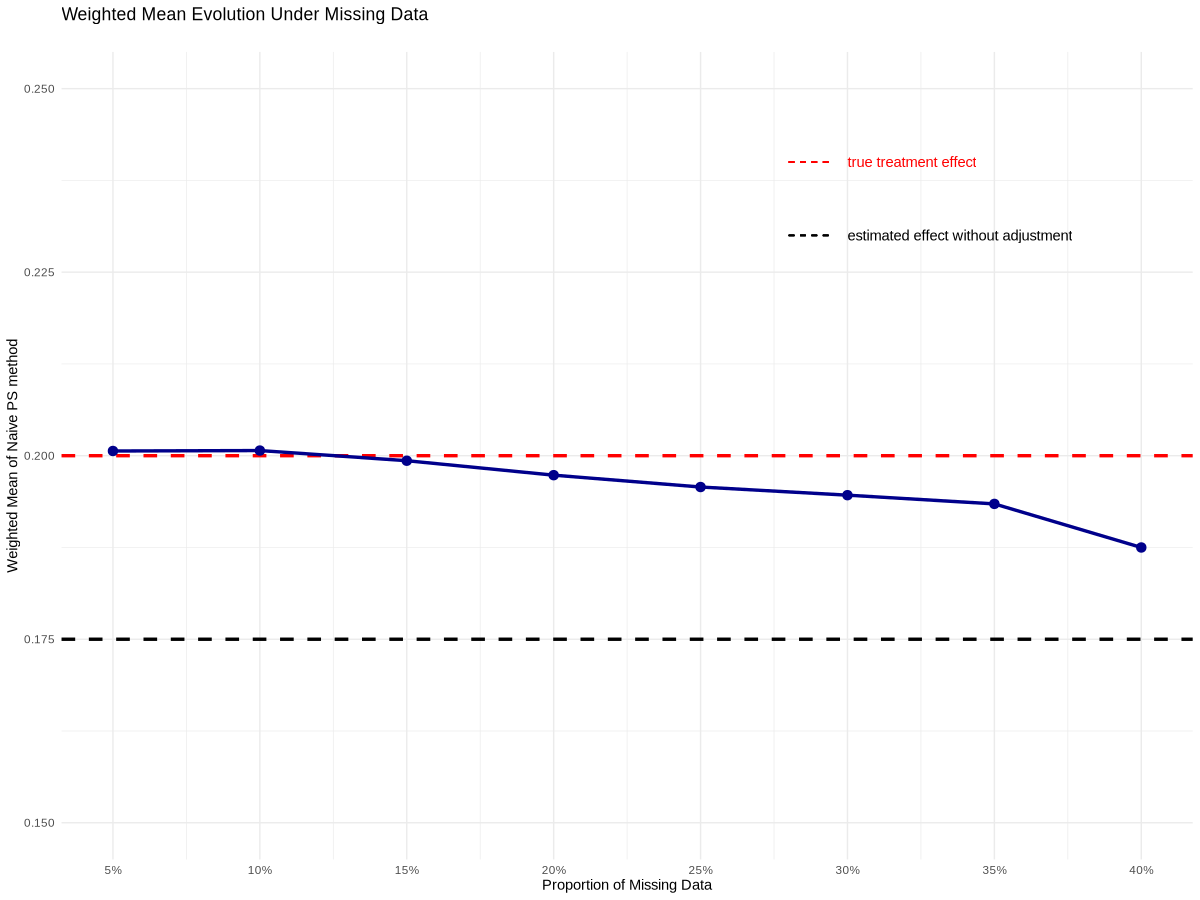}}
    \caption{Estimated treatment difference from post stratification}
    \label{ex3_2}
\end{figure*}

Figure~\ref{ex3_2} further illustrates the impact of incomplete baseline data on the post-stratified mean. The red dashed line indicates the true treatment effect, while the black dashed line represents the average estimated effect based on the unadjusted IA data.  As the proportion of missing baseline data increases,  it is noteworthy that the average of naive post-stratified mean remains well-aligned with the true treatment difference. This observation highlights the robustness of post-stratification in preserving unbiased estimation, even under partial information.

The intuition behind this phenomenon is as follows: when the IA dataset exhibits a deviation from the planned subgroup composition, the distribution observed in the full set of available baseline data tends to compensate for this deviation and move closer to the true population structure. As a result, post-stratification using estimated proportions from available baseline data at IA can still provide meaningful correction and improve decision accuracy.

\section{Permutation test}\label{sec5}
Including categorical variables that do not contribute to population heterogeneity in the post stratification procedure can lead to unnecessary stratification, resulting in an increased number of strata and potentially unstable estimates. To mitigate this issue, we propose a variable screening step prior to post-stratification, aimed at excluding categorical variables that do not meaningfully contribute to population heterogeneity.

To identify such variables, we introduce a nonparametric permutation-based test. Traditional chi-square tests are not appropriate in this context because the IA dataset is not independent of the full baseline dataset available at IA, rendering standard p-values invalid. Instead, we repeatedly draw subsamples from the full baseline dataset, each with the same sample size as the IA dataset, and compute the chi-square test statistic for each subsample. This process yields an empirical null distribution of the test statistic under the assumption of no population shift.

The p-value for the observed test statistic from the IA dataset is then determined by its position within this empirical distribution. This approach enables a valid assessment of variable-level contributions to population heterogeneity without relying on parametric assumptions, and ensures that only informative variables are included in the post-stratification procedure.

\section{Conclusion}\label{sec6}
This paper examined the robustness of futility analyses in the presence of population shifts at IA, a scenario that challenges the  assumption of data MCAR. Through illustrative examples, we demonstrated that deviations in the IA population from the target trial population can introduce extrinsic error, potentially distorting early stopping decisions.

To address this, we explored post-stratification techniques—both naive and model-based—as corrective strategies. While naive post-stratification offers simplicity and interpretability, it is vulnerable to instability in sparse strata. Model-based approaches, particularly Bayesian multilevel regression, provide a principled framework for borrowing strength across groups but are sensitive to model misspecification. To balance these trade-offs, we proposed a hybrid post-stratification method that adaptively combines empirical and model-based estimates, enhancing robustness without sacrificing flexibility.

We further introduced a permutation-based diagnostic to identify variables contributing to population heterogeneity, offering a nonparametric alternative to traditional chi-square tests in dependent data settings. Finally, we emphasized the importance of simulation studies to evaluate the performance of futility rules under realistic data constraints, including incomplete baseline information.

Taken together, our findings underscore the need for careful consideration of population representativeness in interim analyses and advocate for the integration of post-stratification adjustments using all available baseline data at IA to preserve the integrity of trial decision-making.

%\bibliography{wileyNJD-Harvard.bib}
\printbibliography
\end{document}